# On-surface synthesis of polyazulene with 2,6-connectivity

Qiang Sun[1†], Ian Cheng-Yi Hou[2†], Kristjan Eimre[1], Carlo A. Pignedoli[1], Pascal Ruffieux[1], Akimitsu Narita[2,3,*], Roman Fasel[1,4,*]

[1]Empa, Swiss Federal Laboratories for Materials Science and Technology, 8600 Dübendorf, Switzerland

[2]Max Planck Institute for Polymer Research, 55128 Mainz, Germany

[3]Organic and Carbon Nanomaterials Unit, Okinawa Institute of Science and Technology Graduate University, 1919-1 Tancha, Onna-son, Kunigami, Okinawa 904-0495, Japan

[4]Department of Chemistry and Biochemistry, University of Bern, 3012 Bern, Switzerland

*Corresponding Authors: narita@mpip-mainz.mpg.de, roman.fasel@empa.ch
[†]These authors contributed equally to this work.

**ABSTRACT:** Azulene, the smallest neutral non-alternant aromatic hydrocarbon, serves not only as a prototype for fundamental studies but also a versatile building block for functional materials because of its unique (opto)electronic properties. In this work, we report the on-surface synthesis and characterization of the homopolymer of azulene connected exclusively at the 2,6-positions, namely 2,6-polyazulene, using 2,6-diiodoazulene as the monomer precursor. As an intermediate to the formation of polyazulene, a gold-(2,6-azulenylene) chain is observed. The structural details of the resulting 2,6-polyazulene are resolved by high-resolution scanning probe methods, and the electronic properties characterized by scanning tunneling spectroscopy in combination with density functional theory calculations, revealing n-type semiconducting character. Our results provide a route toward the synthesis of novel azulene-based nanostructures, of fundamental interest but difficult to be achieved by conventional solution chemistry.

Properties of carbon-based aromatic systems are sensitively determined by their bond topologies.[1,2] So far, much attention has been paid to the design and synthesis of aromatic materials like conjugated polymers and nanographenes constituted by alternant hydrocarbons, which do not possess odd-membered rings. In contrast, incorporation of non-alternant hydrocarbons has only rarely been explored. Electronic and optical properties of alternant and non-alternant hydrocarbons differ profoundly.[3] Azulene (Scheme 1), for example, an aromatic hydrocarbon containing 10 π-electrons, has several characteristics that differ from its isomer naphthalene.[4,5] Azulene has an intrinsic dipole moment of 1.08 D,[6] while naphthalene is non-polar. The dipole moment of azulene arises from an unequal distribution of electron density between its electron-deficient 7-membered ring and electron-rich 5-membered ring due to an aromatic stabilization according to Hückel's rule. In addition, azulene exhibits an "anomalous" fluorescence from its second excited singlet state in violation of Kasha's rule[7], which makes it a promising candidate in optoelectronics.[8,9]

Because of its unique electronic and optical properties, azulene has been employed as core structure of functional materials for different applications such as stimuli-responsive materials,[10] organic field-effect transistors,[11,12] solar cells[13] and others.[14,15] The connectivity of azulenylene units (Scheme 1a) in the derived structures has a substantial influence on their optical and electronic properties.[16,17] In particular, a 2,6-azulenylene has been shown to exhibit the highest π-electron delocalization among other azulenylenes with different connectivities.[18,19] One remarkable example is that the lowest unoccupied molecular orbital (LUMO) of 2,6':2',6"-terazulene is fully delocalized over the whole molecule, showing strong bonding between the azulene moieties, resulting in a good n-type semiconductor performance.[20] This motivates the study of 2,6-polyazulene . However, despite a number of theoretical studies on 2,6-polyazulenes, reports on the synthesis of azulene-based polymers predominantly focus on the incorporation of 1,3-azulenylenes.[21–24] Recent studies showed that integration of 2,6-azulenylenes into copolymers had great potential for field-effect transistors and proton-conducting materials.[12,25] Nevertheless, the synthesis of azulene homopolymers with 2,6-connectivity by conventional solution chemistry has remained elusive. This might be related to the fact that 2,6-polyazulene is expected to be a rigid linear polymer with a strong aggregation tendency and poor solubility, which is already observed for 2,6':2',6"-terazulene.[20]

On the other hand, on-surface synthesis has recently developed as a new, complementary strategy for chemical synthesis. A significant number of structures and materials which are challenging by conventional solution chemistry, such as pristine zigzag graphene nanoribbons or higher acenes, have been successfully achieved by this method.[26–31] On-surface synthesis relies on well-designed molecular precursors synthesized using the techniques of the solution chemistry. Due to its potentially interesting properties, 2,6-polyazulene is starting to gain attention from the community of on-surface synthesis.[32] Here, we report the successful on-surface synthesis of polyazulene with exclusive 2,6-connectivity using 2,6-diiodoazulene as the molecular precursor (Scheme 1b). Polymerization is triggered and accomplished by heating 2,6-diiodoazulene at a temperature of 580 K on a Au(111) surface. We also find 2,6-azulenylene-gold organo-metallic chains at an intermediate temperature of

350 K. The electronic property of the synthesized polyazulene is investigated by scanning tunneling spectroscopy (STS) and density functional theory (DFT) calculations. The electronic gap of the polyazulene is determined to be 1.8 eV on Au(111), which is further supported by differential conductance dI/dV mapping at the onsets of valence band (VB) and conduction band (CB). This work demonstrates not only the synthesis but also the first detailed electronic characterization of azulene homopolymers with pure 2,6-connectivity, which paves the way to fundamental studies of novel π-conjugated azulene-based carbon materials.

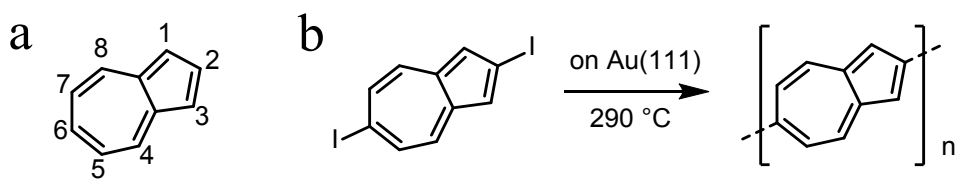

**Scheme 1.** (a) The chemical structure of azulene with carbon atom numbering. (b) On-surface synthetic route toward the 2,6-polyazulene.

Since it is well established that dehalogenative carbon-carbon coupling proceeds readily on metal surfaces, we choose 2,6-diiodoazulene as the precursor for the formation of 2,6-polyazulene. The 2,6-diiodoazulene was synthesized following a previous work[33]. To access the pristine 2,6-diiodoazulene molecules, we first deposit the molecule under ultrahigh vacuum condition onto an Au(111) surface maintained at 160 K to avoid possible deiodination that would occur at room temperature.[34] As shown in the scanning tunneling microscopy (STM) images in Fig. 1a and 1b, 2,6-diiodoazulene forms porous structures on the Au(111) surface. A zoom-in image reveals the monomers appearing as rod structures with brighter iodine protrusions at both ends according to the chemical model overlaid in Fig. 1b. The 2,6-diiodoazulene molecules interact with each other through intermolecular halogen bonding.[35,36] Note that it is not possible to differentiate the 7- and 5-membered rings and the exact orientation of azulenes solely based on the STM images.

After annealing the sample to 350 K, a chain structure is observed on the surface (Fig. 1c and 1d). The molecular chains are composed of two alternating subunits as seen in Fig. 1d, which are assigned to 2,6-azulenylene moieties and to gold adatoms, respectively. We superimpose an equally scaled chemical structure of a chain segment onto a small-scale STM image in Fig. 1d, which shows a good match with the STM image. It is not surprising that gold atoms are involved in nanostructures obtained upon dehalogenation of precursor molecules on Au substrates, and there are many reports of on-surface synthesis of organo-metallic species on Au(111).[37,38] Note that the detached iodine atoms are detected as round protrusions staying aside the organo-metallic chains which is indicated by a blue arrow in Fig. 1d. This is consistent with the fact that iodine desorption from Au(111) starts only at around 540 K.[34] Closer inspection of the structures formed at 350 K reveals that a few 2,6-connected oligoazulenes without the involvement of Au atoms have already been formed (one of them is indicated by a green arrow in Fig. 1d). Further annealing the sample to 580 K triggers the release of all Au atoms from the organo-metallic chains and the formation of the carbon-carbon bonded polymer, as well as partial desorption of iodines (Fig. 1e and 1f). A high-resolution STM image of the polymer reveals that the azulene units are connected along their long axis, implying the formation of 2,6-connected polyazulene.

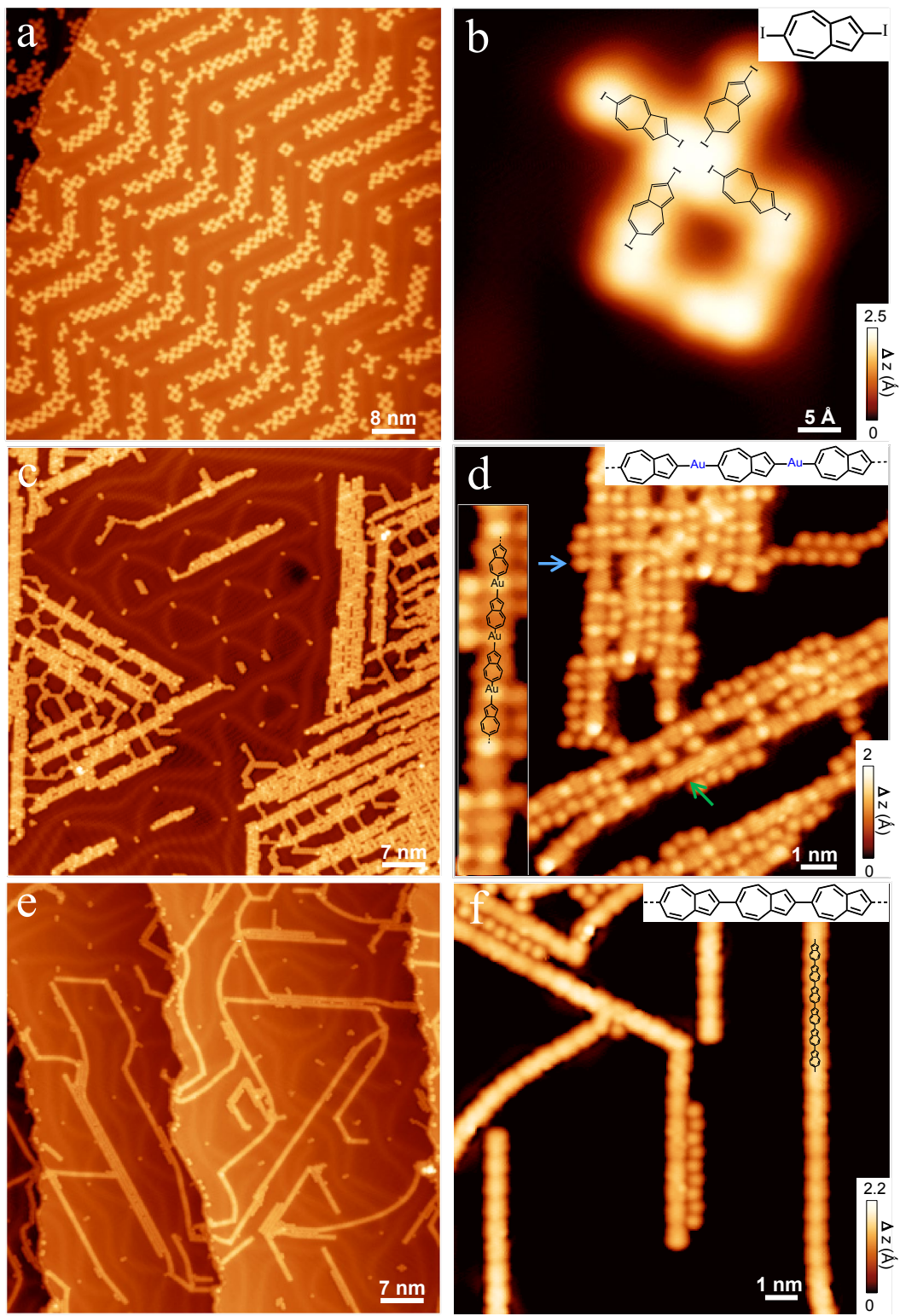

**Figure 1** (a,b) STM images after deposition of 2,6-diiodoazulene on Au(111) held at 160 K. Chemical structures of four molecules are overlaid on the corresponding STM image in (b). (c,d) STM images of the sample after annealing to 350 K. The inset in (d) highlights the organo-metallic chain. The blue arrow indicates a detached iodine atom, the green arrow points out oligomeric 2,6-connected azulene segments. (e,f) STM images after annealing the sample to 580 K. A scaled chemical model of a 2,6-oligoazulene is overlaid on a chain in (f). Scanning parameters: (a) $V_s$ = -1 V, $I_t$ = 0.09 nA; (b) $V_s$ = -0.02 V, $I_t$ = 0.25 nA; (c) $V_s$ = -1 V, $I_t$ = 0.06 nA; (d) $V_s$ = -0.005 V, $I_t$ = 0.1 nA; (e) $V_s$ = -1 V, $I_t$ = 0.08 nA; (f) $V_s$ = -0.01 V, $I_t$ = 1 nA.

To further support the formation of polyazulene and determine the connectivity between azulene units, we resort to bond-resolved non-contact atomic force microscopy (nc-AFM).[39] As shown in the STM image and its corresponding nc-AFM image in Fig. 2a, the structure of the polymer can be clearly resolved, with the 5- and 7-membered rings being imaged with different sizes. Typical defects in the straight 2,6-connected polyazulene are kinks (cf. Fig. S1) arising from 1,6-connected azulene units, which are attributed to a small amount of impurity from the precursor[33] or to a 2,1-sigatropic rearrangement. Within a 2,6-polyazulene chain, neighboring 2,6-azulenylenes can be connected in three different ways, namely pentagon *vs.* heptagon (p-h), pentagon *vs.* pentagon (p-p), or heptagon *vs.* heptagon (h-h), which correspond to 2,6-, 2,2- and 6,6-connectivity, respectively (see the arrows in Fig. 2a). We note that the h-h-connectivity is the most easily recognized, since it results in a non-zero dihedral angle between neighboring azulenylene units which is clearly reflected in the nc-AFM images. Close inspection of high resolution images however also allows unequivocal identification of p-h- and p-p-connectivity. An interesting aspect about the reaction between azulenylenes is whether there is any preference for the p-h-connectivity, since azulene has an intrinsic dipole moment of 1.08 D which might direct the polymerization process *via* dipole-dipole interactions. To this end, we have performed a statistical analysis of the linkages between neighboring azulenylenes. As shown in Fig. 2b, it turns out that the ratio between p-h-, p-p- and h-h-linkages is 2 : 1 : 1, which indicates a non-selective, random connection between the 5- and 7-membered rings (see discussion in Fig. S2). Thus, although the pristine azulene has an intrinsic dipole moment, it does not play a role in orienting the azulenylenes during on-surface polymerization. We attribute this to the screening effect of the electrons of the metal which produces an image dipole, with the resulting dipole vanishing for molecular dipoles oriented parallel to the surface – which is the case for azulene on Au(111).

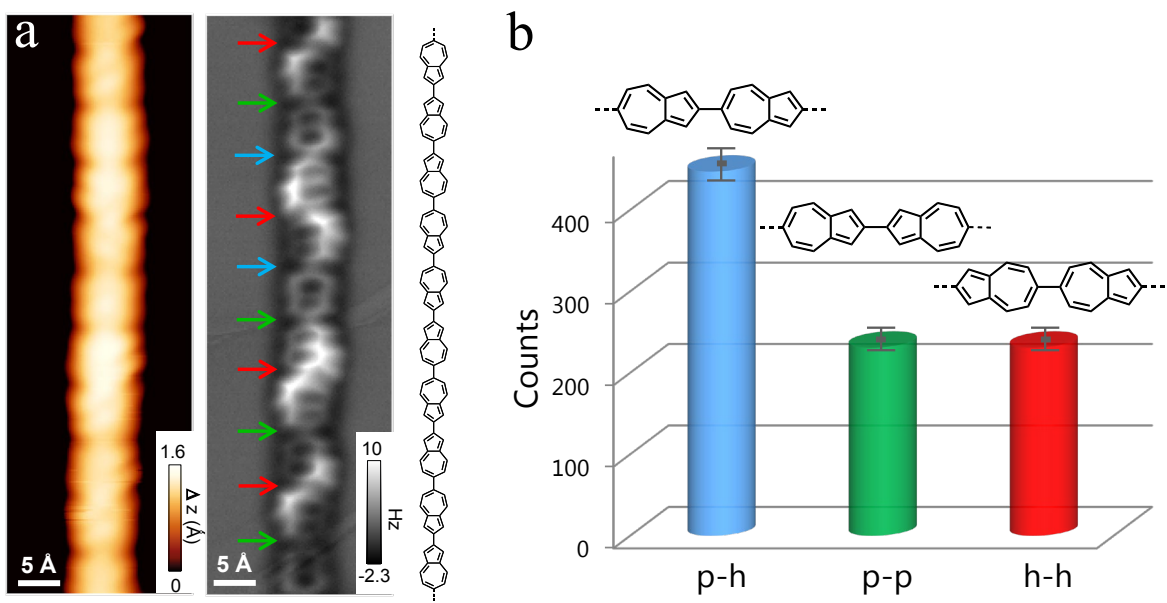

**Figure 2** (a) High-resolution STM image of a 2,6-polyazulene chain ($V_s$ = -0.1 V, $I_t$ = 0.3 nA) and the corresponding nc-AFM image ($V_s$ = 5 mV, Oscillation amplitude: ~80 pm). The pentagon-heptagon (p-h), pentagon-pentagon (p-p) and heptagon-heptagon (h-h) linkages are indicated by blue, green and red arrows, respectively. The corresponding chemical model of the chain is also shown. (b) Statistical analysis of the different linkages within the polymers.

To study the electronic properties of 2,6-polyazulene, we first carry out density functional theory (DFT) calculations of the isolated azulene molecule and of polyazulene. Figure 3a shows the frontier orbitals from HOMO-2 to LUMO+1 of azulene, while Figure 3b gives the band structure of the 2,6-polyazulene originating from these orbitals. The bands originating from the highest occupied molecular orbital (HOMO), HOMO-2 and LUMO+1 are relatively flat due to the orbitals having low amplitudes at the positions where azulene units are connected within the polymer. In contrast, the bands originating from HOMO-1 and LUMO show significant dispersion due to considerable orbital overlap between the neighboring azulenylenes. To further investigate how the random distribution of p-h-, p-p- and h-h-linkages affects the electronic structure of the polymer, we performed band structure calculations for three different structures consisting of a supercell with four azulene units. As presented in Figure 3c, the first structure with only the p-h-linkage between all the azulenylenes reproduces the band structure of 2,6-polyazulene shown in Fig. 3b, but folded to a four times smaller Brillouin zone. The second structure (star) in Fig. 3c has alternating p-p- and h-h-linkages, and the third structure (diamond) has every third unit flipped compared to the others. Apart from a splitting of the bands due to lifting of degeneracy near k-vectors corresponding to the modified periodicity, the band structures near the valence and conduction band onsets do not differ much from each other. All of the three structures modeled have an electronic band gap of 0.94 ± 0.03 eV, and their frontier bands display very similar dispersions. This is in line with the fact that the molecular orbitals of azulene possess no weight at 2,6-positions for HOMO-2 and HOMO but considerable weight for HOMO-1 and LUMO (Figure 3a). The theoretical observations indicate that, although three different linkages are randomly distributed within our experimentally obtained polymers, the electronic properties of the synthesized polyazulene shall remain very similar to the ones of the perfectly regular poly(2,6-azulenylene).

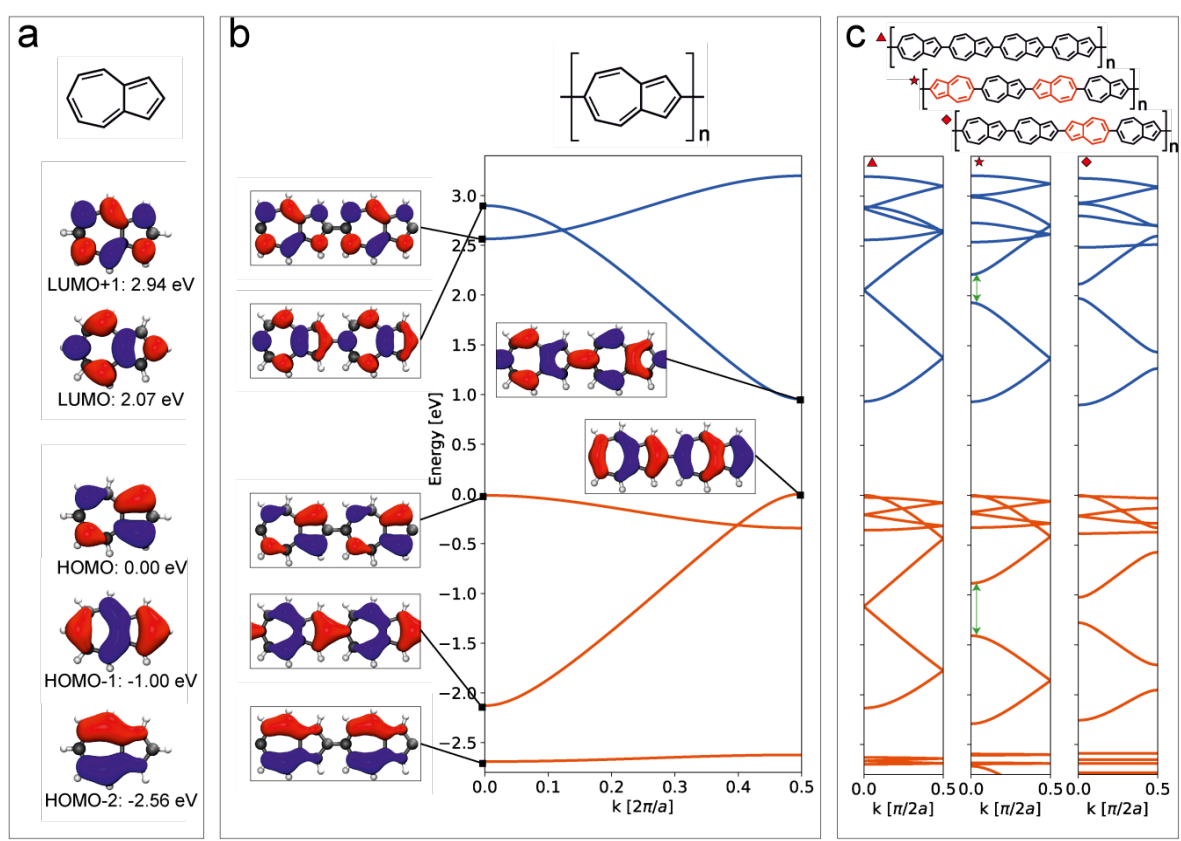

**Figure 3.** Electronic structure of azulene and polyazulenes with 2,6-connectivity. (a) The shape of frontier orbitals of an azulene molecule. (b) Band structure of 2,6-polyazulene and the orbital shapes of the frontier bands which evolve from the corresponding molecular orbital of azulene shown in (a). (c) Band structures of three polyazulenes with different connectivity patterns in a supercell containing four 2,6-azulenylenes with all units oriented in the same way (triangle); first and third unit flipped (star); and only third unit flipped (rhombus). The green arrows indicate two examples of degeneracy liftings caused by the flipped units. In all band structure plots, top of the valence band is taken as zero energy,

orange lines are occupied bands while blue lines correspond to unoccupied bands. In all orbital plots wavefunction isosurfaces at ±0.03 Å$^{-3/2}$ are shown. *a* is the length of one azulene unit within the homopolymer.

To experimentally characterize the electronic properties of the synthesized polyazulene, we have used differential conductance dI/dV spectroscopy. As indicated in Fig. 4a and 4b, the point spectra of the polymer reveal two peaks at around -1.25 V and 0.6 V, which correspond to its VB and CB onsets, respectively. To verify the assignment of the molecular orbitals, we have performed  dI/dV mapping at two corresponding bias voltages (Fig. 4c). The dI/dV maps show clear patterns confirming their origin from molecular orbitals. Moreover, we notice that the position of the h-h-linkage has a higher contrast in dI/dV mapping of the CB onset, which is highlighted by the white rectangle in Fig. 4. This feature could also be seen in the constant-current STM imaging at a bias voltage of 0.6 V (Fig. S3), and is consistent with the fact that the LUMO of an azulene monomer is mainly located at the 7-membered ring (see also Fig. 3a).[22] To further support our experimental findings, we determine the electronic properties of the experimentally observed polyazulene segment shown in Fig. 4a by DFT calculations (Fig. S4). The DFT calculated local density of states (LDOS) maps at the VB and CB onsets are in good agreement with the experimental dI/dV maps at both negative and positive biases (Fig. 4c), and thus confirm the peaks at -1.25 V and 0.6 V in Fig. 4b originating from the VB and CB onsets. Note that the energy positions of the DFT computed LDOS are determined from the calculated DOS shown in Fig. S4. Notably, the h-h-linkage displays characteristic contrast in both DFT calculated maps, and appears particularly bright in the CB (highlighted by dashed rectangles).

The frontier states (i.e. the first occupied and unoccupied states) of the polymer detected by STS are positioned symmetrically with respect to a *negative* bias voltage of around -0.3 V (Fig. 4b). In contrast, the frontier states of most carbon nanowires/ribbons composed of alternant (poly)benzenoids are positioned symmetrically to positive bias voltages on Au(111).[40,41] This is due to the systems having different valence and conduction band alignments with respect to the vacuum level, which – for weakly interacting systems –  determines the corresponding positioning with respect to the Au Fermi level upon adsorption. This is reflected in our DFT calculations, which demonstrate that for polyazulene the VB and CB onsets are respectively at -5.0 eV and -4.1 eV with respect to the vacuum level. In contrast, the band onsets for armchair graphene nanoribbons of width 7 (7AGNR) are found at -4.7 eV and -3.2 eV (Fig. S5). Since the energy-level alignment at organic-metal interfaces plays a vital role in the performance of organic electronics,[42] polyazulene with 2,6-connectivity may serve as a useful n-type organic electronic material.

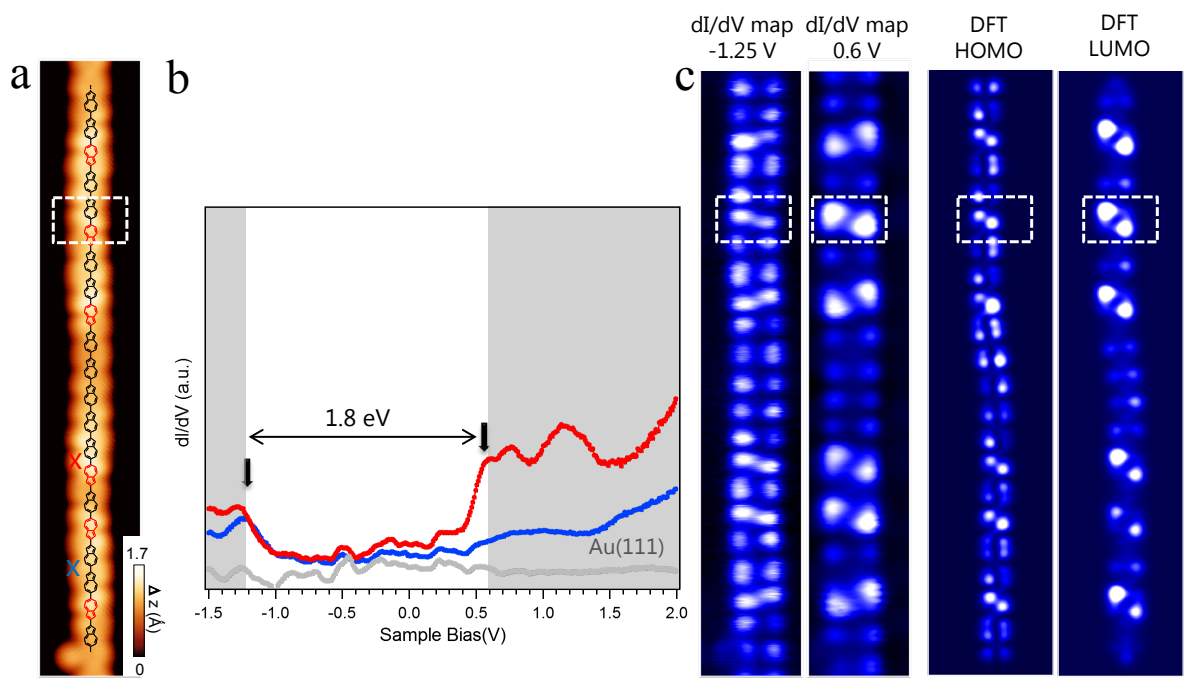

**Figure 4.** (a) STM image of a polyazulene segment ($V_s$ = -1.5 V, $I_t$ = 0.9 nA) and (b) its differential conductance dI/dV spectra. The spectra were taken at the locations marked by crosses with corresponding colors in (a). The grey spectrum has been acquired on a clean Au(111) area nearby. Spectra are vertically offset for clarity. (c) Constant-current dI/dV maps of the HOMO and LUMO, and the corresponding DFT-calculated LDOS maps. The h-h-linkage which shows a higher contrast is highlighted by dashed rectangles in both dI/dV and LDOS maps.

In conclusion, the homopolymer of azulene with exclusive 2,6-connectivity has been synthesized by dehalogenative coupling of 2,6-diiodoazulene on a Au(111) surface. Although azulene has an intrinsic dipole moment, dipole-dipole interactions do not yield a preferential pentagon-heptagon-linkage along the polymer. However, it turns out that the electronic properties of 2,6-polyazulene do not significantly depend on the ratio of the pentagon-pentagon, pentagon-heptagon and heptagon-heptagon linkages. Scanning tunneling spectroscopy yields an electronic gap of 1.8 eV for polyazulene on the Au(111) surface, which is confirmed by STS mapping and DFT-based LDOS simulations. The present study has further demonstrated the potential of a combination of traditional solution chemistry and emerging on-surface synthesis in fabricating novel carbon-based nanomaterials. The reported synthesis route may also enable future applications of azulene-based functional materials.

**Methods**

**STM/STS/nc-AFM.** A commercial low-temperature STM/nc-AFM (Scienta Omicron) system was used for sample preparation and *in situ* characterization under ultra-high vacuum conditions (base pressure below $1\times10^{-10}$ mbar). The Au(111) single crystal was cleaned by argon sputtering ($p = 6 \times 10^{-6}$ mbar) and annealing cycles to 750 K for 15 minutes. Deposition of the molecular precursors was done by thermal evaporation from a 6-fold organic evaporator (Mantis GmbH). STM images were recorded in constant-current mode, and the dI/dV spectra were recorded using the lock-in technique ($U_{RMS}$ = 20 mV). nc-AFM images were recorded with a CO-functionalized tip attached to a quartz tuning fork sensor (resonance frequency 23.5 kHz, peak-to-peak oscillation amplitude below 100 pm).

**Density functional theory calculations.** DFT orbitals of the isolated azulene and the band structures were calculated with the Quantum Espresso software package using the PBE exchange correlation functional. A plane wave basis with an energy cutoff of 400 Ry for the charge density was used together with PAW pseudopotentials (SSSP[43]). For the band structure of the primitive (super) cell calculations a Monkhorst k-mesh of 40 x 1 x 1 (10 x 1 x 1) was used. The models of the gap-phase calculated polyazulenes were constructed in flat geometries. The cell and atomic geometries were relaxed until forces were smaller than 1 e$^{-4}$ a.u. To perform these calculations we used the AiiDA platform.[44]

The equilibrium geometry of the polyazulene polymer on the Au(111) substrate was obtained with the CP2K code.[45] The gold slab consisted of 4 atomic layers of Au along the [111] direction and a layer of hydrogen to suppress one of the two Au(111) surface states. 40 Å of vacuum were included in the cell to decouple the system from its periodic replicas in the direction perpendicular to the surface. We used the TZV2P Gaussian basis set[46] for C and H and the DZVP basis set for Au together with a cutoff of 600 Ry for the plane wave basis set. We used norm conserving Goedecker-Teter-Hutter[47] pseudopotentials, the PBE[48] parametrization for the exchange correlation functional and the Grimme's DFT-D3 dispersion corrections[49]. To obtain the equilibrium geometries we kept the atomic positions of the bottom two layers of the slab and hydrogen layer fixed to the ideal bulk positions, all other atoms were relaxed until forces were lower than 0.005 eV/Å.

**Acknowledgements**

This work was supported by the Swiss National Science Foundation under Grant No. 200020_182015, the NCCR MARVEL funded by the Swiss National Science Foundation (51NF40-182892), the European Union's Horizon 2020 research and innovation programme under grant agreement number 785219 (Graphene Flagship Core 2), the Office of Naval Research (N00014-18-1-2708), and a grant from the Swiss National Supercomputing Centre (CSCS) under project ID s904. Q.S. acknowledges the EMPAPOSTDOCS-II programme under the Marie Sklodowska-Curie grant agreement No 754364.

**Competing financial interests**

The authors declare no competing financial interests.